%% file: main.tex
\documentclass{ieeeaccess}
\input{settings}

\begin{document}

\history{Date of publication xxxx 00, 0000, date of current version xxxx 00, 0000.}
\doi{10.1109/ACCESS.2020.2988889}

\title
{SMART: A \underline{S}ecure \underline{M}agnetoelectric \underline{A}ntife\underline{R}romagnet-Based \underline{T}amper-Proof Non-Volatile Memory}

\author{\uppercase{Nikhil Rangarajan}\authorrefmark{1},
\IEEEmembership{Member, IEEE},
\uppercase{Satwik Patnaik}\authorrefmark{2},
\IEEEmembership{Student Member, IEEE},
\uppercase{Johann Knechtel}\authorrefmark{1},
\IEEEmembership{Member, IEEE},
\uppercase{Ozgur Sinanoglu}\authorrefmark{1},
\IEEEmembership{Senior Member, IEEE},
and
\uppercase{Shaloo Rakheja}\authorrefmark{3},
\IEEEmembership{Member, IEEE}}
\address[1]{Division of Engineering, New York University Abu
Dhabi (NYU AD), Abu Dhabi, 129188, UAE (e-mails: \{nikhil.rangarajan, johann, ozgursin\}@nyu.edu)}
\address[2]{Department of Electrical and 
Computer Engineering, 
New York University (NYU), Brooklyn, NY, 11201, USA (e-mail: sp4012@nyu.edu)}
\address[3]{Holonyak Micro and Nanotechnology Laboratory, University of Illinois at Urbana-Champaign (UIUC), Urbana, IL, 61801, USA (e-mail: rakheja@illinois.edu)}
\tfootnote{This work was supported by the Semiconductor
Research Corporation (SRC) and the National Science Foundation
(NSF) through ECCS 1740136.
The work of Satwik Patnaik was supported in part by the Global Ph.D. Fellowship at NYU/NYU AD.}

\markboth
{Rangarajan \headeretal: SMART: A \underline{S}ecure \underline{M}agnetoelectric \underline{A}ntife\underline{R}romagnet-Based \underline{T}amper-Proof Non-Volatile Memory}
{Rangarajan \headeretal: SMART: A \underline{S}ecure \underline{M}agnetoelectric \underline{A}ntife\underline{R}romagnet-Based \underline{T}amper-Proof Non-Volatile Memory}

\corresp{Corresponding authors: Nikhil Rangarajan, Satwik Patnaik, and Shaloo Rakheja (e-mails: nikhil.rangarajan@nyu.edu, sp4012@nyu.edu, and rakheja@illinois.edu).}

\begin{abstract}

The storage industry is moving toward emerging non-volatile memories (NVMs), including the spin-transfer torque magnetoresistive random-access memory (STT-MRAM) and the phase-change memory (PCM), owing to their high density and low-power operation. 
In this paper, we demonstrate, for the first time, circuit models and performance benchmarking for the domain wall (DW) reversal-based magnetoelectric-antiferromagnetic random access memory (ME-AFMRAM) at cell-level and at array-level. 
We also provide perspectives for coherent rotation-based memory switching with topological insulator-driven anomalous Hall read-out. 
In the coherent rotation regime, the ultra-low power magnetoelectric switching coupled with the terahertz-range antiferromagnetic dynamics result in substantially lower energy-per-bit and latency metrics for the ME-AFMRAM compared to other NVMs including STT-MRAM and PCM. 
After characterizing the novel ME-AFMRAM, we leverage its unique properties to build a dense, on-chip, secure NVM platform, called \textit{SMART: A Secure Magnetoelectric Antiferromagnet-Based Tamper-Proof Non-Volatile Memory}. 
New NVM technologies open up challenges and opportunities from a data-security perspective. 
For example, their sensitivity to magnetic fields and temperature fluctuations, and their data remanence after power-down make NVMs vulnerable to data theft and tampering attacks. 
The proposed SMART memory is not only resilient against data confidentiality attacks seeking to leak sensitive information but also ensures data integrity and prevents Denial-of-Service (DoS) attacks on the memory.
It is impervious to particular power side-channel (PSC) attacks which exploit asymmetric read/write signatures for `0' and `1' logic levels, and photonic side-channel attacks which monitor photo-emission signatures from the chip backside.

\end{abstract}


\titlepgskip=-25pt

\maketitle

\section{Introduction and Background}
\label{sec:introduction}

Conventional dynamic random-access memory (DRAM)
scaling has reached a critical tipping point as the miniaturization of the DRAM cell has plateaued in recent years. Feature size scaling below the 20 $\text{nm}$ technology node is met with numerous challenges such as 
shorter retention times, higher leakage currents, and increased fault rates~\cite{park2015technology}.
Solutions to address these concerns include improved DRAM fault detection and recovery~\cite{wang2017improving}, as well as architectural techniques to enhance DRAM scaling~\cite{kim2015architectural}. 

A promising solution to the memory scaling problem is to realize the main memory system using non-volatile technologies~\cite{mutlu2015main}. 
Examples of emerging non-volatile memories (NVMs) include spin-transfer torque magnetoresistive random-access memory (STT-MRAM), ferroelectric random-access memory (FeRAM), resistive random-access memory (ReRAM), and phase-change memory (PCM).
Interest in the commercial application of such NVMs has increased significantly. For instance, Intel's current line of 3D XPoint memory systems utilize PCM-based NVM technology~\cite{wyrwas2017proton}, and IBM and Everspin's solid-state drive comes with STT-MRAM write caches~\cite{everspin}. While NVMs offer attractive features, such as high density, low leakage, and non-volatile data retention, they also suffer from poor endurance and high access latency in their current implementation.

Memory security has come under more scrutiny over the years.
This is because of attacks such as \textit{Spectre}~\cite{kocher2019spectre} and \textit{Meltdown}~\cite{lipp2018meltdown}, which targets the side-channels associated with speculative execution and out-of-order execution, respectively, have exposed severe vulnerabilities in a wide array of currently deployed processors and their memory architectures. 
In the case of NVMs, data remanence after power-down
presents a severe threat to data confidentiality, as attackers aiming to steal private data can do so easily by mounting cold-boot attacks~\cite{halderman2009lest} or other removal attacks like stealing the memory module (DIMM)~\cite{young2015deuce}. 
Moreover, magnetic memories like STT-MRAM are highly sensitive to stray magnetic fields. 
As such, magnetic field-based attacks~\cite{jang2015self} can be used to corrupt the stored data or compromise the memory's functional integrity, resulting in a denial-of-service (DoS) attack. 
Hence, such security vulnerabilities pose a significant impediment to the pervasive and large-scale proliferation of NVMs in the memory industry. 

\subsection{Related work in Memory Security}
Prior works on securing NVMs have focused mainly on memory encryption schemes, which are necessary to prevent attackers from exploiting data remanence in the off-state. 
Chhabra \textit{et al.} proposed an incremental encryption scheme~\cite{chhabra2011nvmm} for NVMs where only inert memory pages, which have not been accessed for several clock cycles, are encrypted selectively. 
The working set of the memory (which is in current use) is in plaintext and, hence, incurs no encryption overhead on access. Such a selective encryption ensures that the majority of the main memory content (but not all) remains encrypted at all times, without overly compromising the performance. 
However, it requires dedicated hardware, inert page prediction, and scheduling for its implementation. 
A sneak-path encryption (SPE) scheme was demonstrated for memristor-based NVMs in~\cite{kannan2014secure}, wherein sneak paths in the memristor crossbar array are exploited to apply encryption pulses to change the resistances of the memory cells, and hence, encrypt the stored data. 

In~\cite{young2015deuce}, the authors proposed DEUCE, a dual counter encryption for PCM memories, which significantly reduces the number of modified bits per writeback, to improve performance and lifetime of the memory. 
This scheme aims to mitigate the impact of the avalanche effect~\cite{mandal2012performance} occurring during memory encryption, by re-encrypting and writing back only the modified words during any write operation. Swami~\textit{et al.} took this concept forward and proposed SECRET~\cite{swami2016secret}, a smart encryption scheme for NVMs, which integrates word-level re-encryption and zero-based partial writes to reduce memory write operations. They also demonstrate write optimization through the use of
``energy masks'' (i.e., bit templates XORed with ciphertext to obtain lower energy dissipation)
in the encryption XOR logic, which minimizes the bit flips in the encryption process, thereby reducing the total write energy.
An advanced counter-mode encryption (ACME) was presented in~\cite{swami2018acme}, which utilizes the write leveling architecture inherent in PCM memories, to perform counter-write leveling. 
ACME helps to avoid \textit{Rowhammer}-type attacks by preventing the counter associated with any single cache line from overflowing.

The impact of contactless tampering on STT-MRAMs using external magnetic fields was highlighted in~\cite{jang2015self}. 
Using micromagnetic simulations, the authors of~\cite{jang2015self}
showed how magnetic field-based attacks could corrupt the contents of STT-MRAM cells. Techniques to protect against contactless attacks proposed in~\cite{jang2015self} included (i) an on-chip sensor to detect magnetic field-based incursions, and (ii) error correction modules to compensate cell failures arising due to magnetic field attacks. However, these techniques incur large energy and area penalties due to the additional hardware imposed by the magnetic field sensor and the error correction scheme.

\subsection{Contributions}
In this paper, we present an alternative to conventional NVMs such as STT-MRAM and PCM, in the form of \textit{SMART: A Secure Magnetoelectric Antiferromagnet-Based Tamper-Proof Non-Volatile Memory}. 
SMART memory leverages the room-temperature linear magnetoelectric (ME) effect in antiferromagnets (AFMs) like chromia~\cite{rado1961observation}, which can be switched solely using voltage pulses, without the use of electric currents, leading to ultra-low energy ($\sim$ pico-Joules) operation. 
Further, the intrinsic dynamics of AFMs is typically in the terahertz regime ($\sim 10^{12}$ Hz)~\cite{khymyn2017antiferromagnetic}, which could enable picosecond time-scale reversal of the AFM domain. 
In addition to its energy and latency benefits, SMART memory offers a significant advancement in terms of secure and tamper-proof data storage. 
For example, AFMs do not exhibit a magnetic signature since they do not have a net external magnetic moment, unlike ferromagnets (FM). 
Hence, the SMART memory cannot be probed or switched with external magnetic fields, unlike the way STT-MRAMs can. 
This, in turn, eliminates the possibility of magnetic field attacks undermining data integrity or aiming to induce DoS. 
To address the post-shutdown data remanence of SMART memory, we demonstrate an in-memory encryption scheme employing  ME-AFM transistor-based controlled-NOT (CNOT) logic. 
We discuss the resilience of the SMART memory against attacks aiming to undermine data confidentiality and data fidelity, in both powered-on and powered-off states. 
The main contributions of this work can be summarized as follows:

\begin{enumerate}
    
\item We discuss the design of SMART, a secure ME-AFM-based NVM and implement its SPICE circuit model to simulate the memory performance. 
    
\item We demonstrate the resilience of SMART memory against magnetic field and temperature attacks, which can affect other NVMs like STT-MRAM. We explore the implications of various side-channel attacks on the SMART memory.
    
\item We present an in-memory encryption scheme with ME-AFM transistor-based CNOT gates, called \textit{Memcryption}, to protect the data stored in SMART memory against cold-boot and stolen DIMM attacks, while incurring low encryption latency overheads.
We like to mention here that \textit{Memcryption} is specifically tailored for the ME-AFMRAM, not for a generic NVM. Also, it does not secure the memory system against \textit{bus snooping} attacks; such attacks are beyond the scope of this work.

\end{enumerate}

In the next section, we describe the modeling, implementation and benchmarking of the proposed ME-AFM memory both at cell- and array-level, before proceeding to evaluate its security properties in Section~\ref{sec:security}.

\section{Device model and functionality}
\label{Modeling}
\subsection{The magnetoelectric effect}
The linear ME effect~\cite{agyei1990linear} represents the coupling between applied magnetic field and induced polarization or between applied electric field and induced magnetization in non-centrosymmetric crystals like chromia ($\text{Cr}_2\text{O}_3$). Compared to the STT-based magnetization reversal of FMs requiring electric currents on the order of $\sim10^6$ A/cm$^2$ and incurring associated Joule heating, the ME effect provides an energy-efficient, all-electrical switching of the roughness-insensitive boundary magnetization of chromia~\cite{echtenkamp2013electric}. Additionally, chromia is an AFM; hence, the net bulk magnetic moment
(i.e., the difference of the sublattice magnetization vectors) vanishes and becomes imperceptible externally. 
However, the boundary magnetization is strongly coupled to the AFM order parameter. That is, the electrical switching of the AFM order results in reversal of the boundary magnetization~\cite{wu2011imaging}, which is used to encode the information in ME-AFM memories.

The uncompensated surface moments at the (0001) surface of chromia result in an equilibrium boundary magnetization, which could be in one of the two oppositely aligned,
degenerate domain states. 
The degeneracy between the domains is lifted through ME annealing, which allows the preferential selection of one of the states~\cite{he2010robust}. That is, the ME annealing polarizes the surface and results in
a single-domain surface moment. 
Isothermal switching between these single domain states using an electric field $E$ and a small, symmetry-breaking
DC magnetic field $H$ has been demonstrated experimentally~\cite{he2010robust, fallarino2015magnetic}. 
The critical condition for such ME switching is that the magnitude of the $E\cdot H$ product must exceed the ME threshold energy barrier, which was shown experimentally to be as low as $\approx$ 1 J/m$^3$~\cite{brown1969domain, martin1966antiferromagnetic}.

\subsection{ME-AFMRAM : Working principle}
\begin{figure}[ht]
\centering
\includegraphics[scale=0.28]{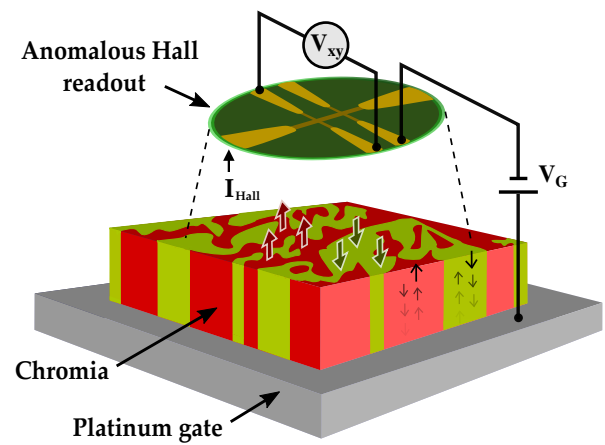}
\caption{Chromia-based magnetoelectric antiferromagnetic random-access memory. Data (1/0) is written by applying a voltage ($+/-$) to the bottom gate electrode. Read-out is achieved using an anomalous Hall bar electrode placed on top, by applying a Hall bias.}
\label{fig:AFMRAM}
\end{figure}

The chromia-based ME-AFMRAM, which is at the heart of our SMART memory, is shown in Fig.~\ref{fig:AFMRAM}. Experimentally demonstrated by Kosub \textit{et al.}~\cite{kosub2017purely}, the ME-AFMRAM has a bottom gate electrode (Platinum gate in the figure) for applying the gate voltage $V_G$ and providing the necessary electric field to write data into the memory. A small, symmetry-breaking magnetic field ($\approx$ 30 mT) is provided by the stray field of a permanent magnet. A positive voltage $V_G$ will orient the bulk order and, hence, put the surface magnetization in one domain (with surface moments pointing up), whereas a negative voltage will result in the surface magnetization relaxing to the opposite domain (with surface moments pointing down). These two states correspond to binary levels `1' ($V_G > 0$) and `0' ($V_G < 0$), respectively. A gate voltage of 0 V corresponds to the `hold' mode of the memory cell. Note that the cell serves as non-volatile memory in all gate-voltage ranges, not only for $V_G = 0$.

The read-out is achieved using an anomalous Hall (AH) bar electrode setup, which discerns the boundary magnetization of chromia by sensing the proximity effect-induced magnetization in the nearby Platinum (Pt) electrode, thereby producing a 
proportional Hall voltage $V_{\text{xy}}$ (or $V_{\text{AHE}}$)~\cite{kosub2015all}. Traditionally, the order parameter of AFMs is read-out via an exchange bias arrangement~\cite{toyoki2015magnetoelectric} in another FM attached adjacently to the AFM surface. However, the exchange bias and the FM's hysteresis increase the coercive voltage required to overcome the ME barrier and, hence, impact the write energy negatively. To avoid this effect, Kosub \textit{et al.}~\cite{kosub2017purely} proposed the use of an exclusively ME-AFM setup with an AH read-out of the surface magnetization, thereby eliminating the need for an FM.
At the time of writing this paper, a complete physical understanding of the read-out mechanism for the boundary magnetization in chromia is lacking. While the authors in~\cite{kosub2017purely} have considered an AH-based read-out in their device, recent experiments by C. Binek's group at the University of Nebraska-Lincoln have revealed the contribution of spin-Hall magnetoresistance (SMR) to the read-out signal, which is currently being investigated.
However, note that the magnitude of the signal levels is the same in both cases (AH versus SMR) and also the circuit models developed would remain the same, though with different input parameters. 
For the purposes of this paper, we consider that the read-out signal is due to the AH effect in the proximal heavy metal, as also discussed in prior experimental work.

\subsection{Performance modeling}
The ME reversal mechanism in chromia can be classified broadly into two categories, depending on the size of the film compared to the characteristic domain-wall (DW) width. For chromia, the typical DW width $\lambda =\sqrt{A/\mathcal{K}}\sim$ 50-100 nm, where $A$ is the exchange stiffness constant and $\mathcal{K}$ is the uniaxial anisotropy energy~\cite{belashchenko2016magnetoelectric}. 
If the sample is much smaller than the DW width, the sample reverses via coherent rotation upon application of the ME pressure. For sample dimension comparable to the DW width, ME reversal occurs via DW nucleation and propagation, which is an incoherent switching process.
For both coherent rotation and DW propagation, the reversal could be thermally activated for applied ME pressure lower than the energy barrier between the stable domain states. Otherwise, the domain reversal proceeds in the `flow' regime~\cite{parthasarathy2019dynamics}.
ME-AFMRAM devices currently fabricated have dimensions in the $\mu$m range, rendering DW propagation the favorable ME reversal mechanism. To characterize the functionality and performance of chromia ME-AFMRAM, we develop circuit models that represent DW-based reversal in both the thermally activated and the flow regimes. We also provide perspectives and future
potential concerning dimensional scaling of the device, which could enable ultra-fast, coherent, rotation-based reversal. 

\subsubsection{DW reversal of chromia ME-AFMRAM}
Consider a chromia sample, where the applied ME pressure creates a pressure difference of $\mathcal{F} =|2\alpha_{\text{ME}} E H|$ between the two domains. Here, $\alpha_\mathrm{ME}$ is the linear ME coefficient.

If $\mathcal{F}> \mathcal{F}_d$ (i.e., for DW de-pinning pressure), the DW propagates as a viscous flow with velocity given as~\cite{parthasarathy2019dynamics}
$$\nu_{\text{flow}} = \frac{\alpha_{\text{G}}\gamma \lambda}{\alpha_+\xi^2}\Big( \frac{\mathcal{F}-\mathcal{F}_{\text{d}}}{M_{\text{s}}} \Big),$$
where $\alpha_{\text{G}}$ is the Gilbert damping constant, $\gamma$ is the gyromagnetic ratio of electron, $M_{\text{s}}$ is the sublattice saturation magnetization, and $\xi=\frac{\alpha_{\text{ME}}E}{\mu_0 M_{\text{s}}}$. 
For a mean free path of $l$ of the DW, the time-scale of ME reversal due to viscous DW propagation is $\tau_{\text{flow}}=l/\nu_{\text{flow}}$.

If $\mathcal{F}<\mathcal{F}_{\text{d}}$, the DW undergoes thermal creep to overcome the de-pinning barrier, with a time-scale~\cite{parthasarathy2019dynamics}
$$\tau_{\text{creep}}=\sqrt{\frac{\sigma \mathcal{S}^3}{kT}}\Big(\frac{\mathcal{F}_{\text{d}}-\mathcal{F}}{2\pi\epsilon}\Big)\exp\Big[{\frac{\mathcal{S}^2(\mathcal{F}_{\text{d}}-\mathcal{F})^2}{4\pi kT\epsilon}\Big]},$$
where $kT$ is the thermal energy (25 meV at 300 K), $\epsilon$, $\sigma$, and $\mathcal{S}$
are the energy, areal density, and surface area, respectively, of the DW. The DW de-pinning pressure is determined by the DW energy, its surface area, and the radius of the non-magnetic de-pinning center.

To write `1' (`0') into the memory cell, a positive (negative) electric field, $E_\mathrm{app}$, with a magnitude greater than the critical electric field, $E_\mathrm{crit}$, is required, in order to meet the DW propagation criteria of $\mathcal{F}>\mathcal{F}_d$. In this case, the time to write data into the memory is equal to $\tau_\mathrm{flow}$. When $E_\mathrm{app}$ is less than $E_\mathrm{crit}$ (i.e., $\mathcal{F}<\mathcal{F}_d$), the memory cell is in the hold mode and the retention time is specified by $\tau_\mathrm{creep}$. For typical parameters of chromia, we find $\tau_\mathrm{creep}\gg \tau_\mathrm{flow}$, which ensures that the memory cell is thermally stable when it is not accessed. Here, the stability of the cell is determined by $\tau_{\text{creep}}$, since longer data retention requires the time constant in the hold mode to be larger. The retention time of the cell can be further improved by enlarging the cell dimensions.

\begin{figure*}[ht]
\centering
\includegraphics[scale=0.19]{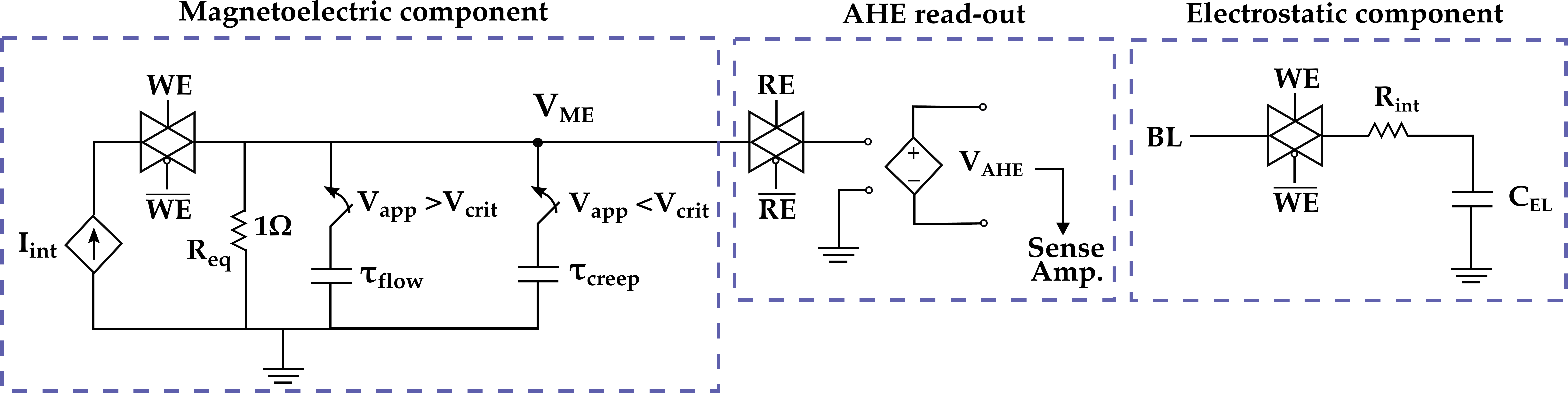}
\caption{Equivalent circuit for the chromia ME-AFMRAM cell. $I_{\text{int}}$, derived from the bit line, writes data on to the node $V_{\text{ME}}$. The time constant of the write operation is $\tau_{\text{flow}}$ ($\tau_{\text{creep}}$) if the applied voltage is greater (smaller) than the critical voltage. Read-out is achieved through an AH setup, modeled with a voltage-controlled voltage source. $\text{C}_\text{EL}$ is the electrostatic capacitance of the chromia dielectric.}
\label{fig:chromia_RC}
\end{figure*}

\begin{figure}[ht]
\centering
\includegraphics[scale=0.37]{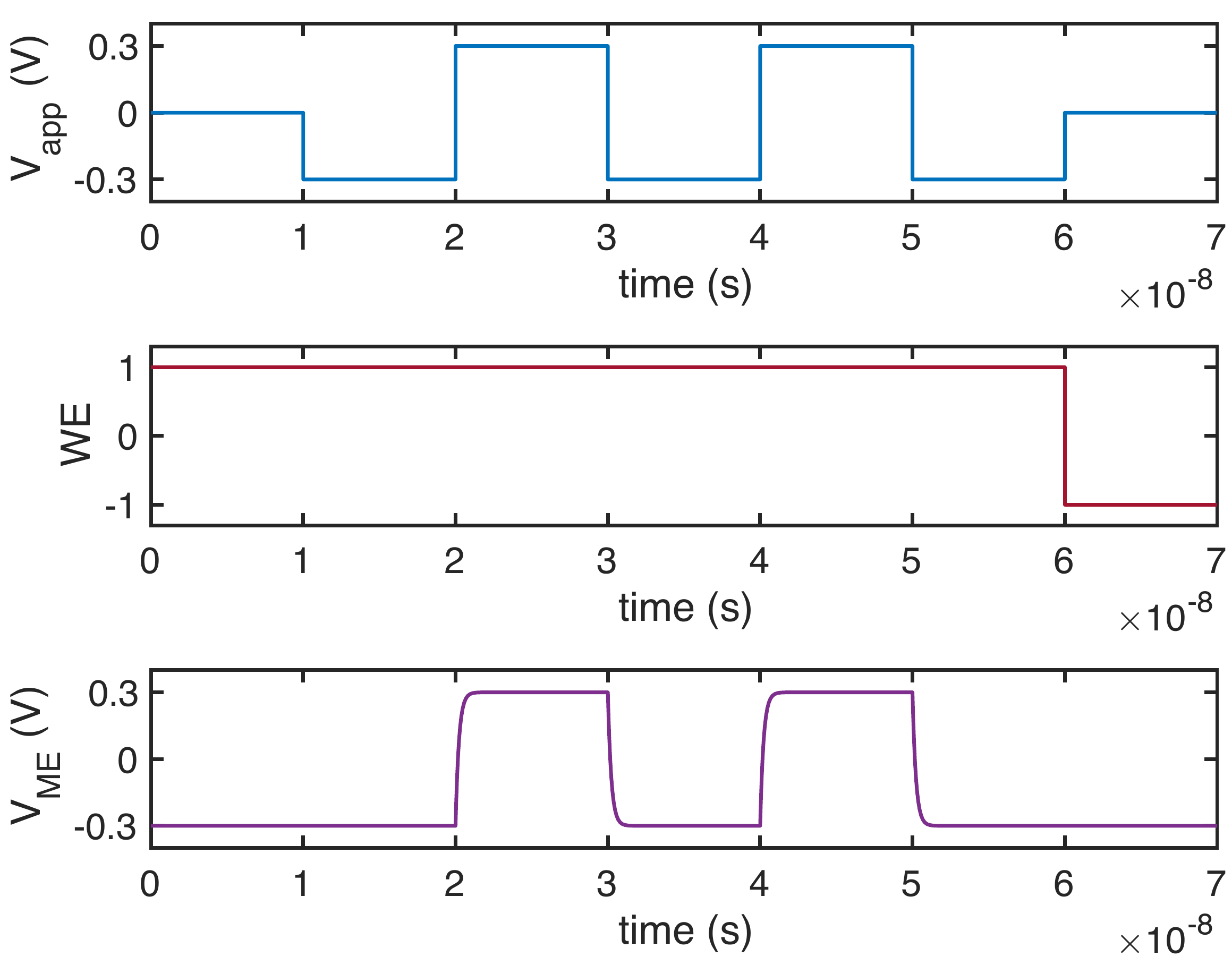}
\caption{Transient simulations showing write operations on the chromia ME-AFMRAM cell. Note that for writing a `1' the write pulse is positive, and for writing a `0' the write pulse is negative. In this simulation, a series of `1's (0.3 V) and `0's (-0.3 V) are being written to the cell, and then finally `0' is retained once Write Enable is switched off.}
\label{fig:AFMRAM_timing}
\end{figure}

We construct a SPICE circuit model to functionally capture the ME reversal dynamics of chromia. The time constant for reversal of the magnetization of chromia due to an applied ME pressure is represented as $R_{\text{eq}}\times C_{\text{eq}}$. Without loss of generality, the circuit model uses $R_{\text{eq}} = 1$ $\Omega$, while $C_{\text{eq}}$ is either $\tau_{\text{flow}}$ or $\tau_{\text{creep}}$. 
To construct the full ME-AFMRAM cell, we combine the RC model of the ME response of chromia with the peripheral read/write circuitry in \textit{Cadence Virtuoso} using the 15-nm CMOS FreePDK technology. Figure~\ref{fig:chromia_RC} shows the equivalent circuit of the ME-AFMRAM cell. 
The write pulse, used to charge the chromia dielectric and switch its
magnetization $M$, is provided through the current source $I_{\text{int}}$ (derived from the bit line) in the write setup. 
For parameters of chromia listed in Table~\ref{tab:params}, 
$C_{\text{flow}}=\tau_{\text{flow}}\sim0.223$ nF, $C_{\text{creep}}=\tau_{\text{creep}}\sim1$ mF, and $V_{\text{crit}}= 0.2$ V. For $|V_G| > 0.2$ V, $V_{\text{ME}}$ tracks $V_G$ and data is written into the cell after a write access latency of $\tau_\mathrm{flow}$. 
When $|V_G| = 0$ V, data is retained for a time interval of $\tau_{\text{creep}}$. Since $\tau_{\text{creep}}$ is very large, the response in retention/creep mode is extremely slow as compared to write/flow mode. The transient response of the ME-AFMRAM cell is shown in Fig.~\ref{fig:AFMRAM_timing}, to highlight the write operation. The write latency of the ME-AFMRAM cell is obtained as $\sim 0.63$ ns, and the energy-per-bit for one write operation is $\sim 0.063$ pJ, including the energy required to charge the electrostatic capacitance of chromia.
Given relative dielectric permittivity of 11 and dimensions noted in Table~\ref{tab:params}, the electrostatic capacitance of chromia is calculated as $5.8$ aF.

\subsubsection{Anomalous Hall read-out}
To evaluate the read cycle, we set the signals WE to 0 and RE to 1 in Fig.~\ref{fig:chromia_RC}. The read setup is designed to sense the boundary magnetization of chromia through an AH arrangement, which transduces the magnetization into a voltage signal. This transduction process is modeled using a voltage-controlled voltage source
(VCVS). Typically, a heavy metal such as Pt is used to sense the proximity effect-induced moment from the coupled chromia layer~\cite{kosub2017purely}.

The AH voltage sensed from the Hall bar arrangement is given as~\cite{griffiths2017anomalous}
$$V_{\text{AHE}}=\Big(\frac{\mu_0R_{\text{s}}}{t_{\text{Hall}}}I_{\text{Hall}}\Big)M_{\text{z}},$$
where $\mu_0$ is the vacuum permeability, $R_{\text{s}}$ is the AH coefficient, $I_{\text{Hall}}$ is the Hall bias current, $t_{\text{Hall}}$ is the thickness of the Hall layer and $M_{\text{z}}$ is the proximity effect-induced magnetization. In the case of Pt/Cr$_2$O$_3$, $R_{\text{s}}$ is only about $\sim5$ p$\Omega$m/T for $t_{\text{Pt}}=10$ nm and $T=300$ K~\cite{meyer2015anomalous}. This results in an AH signal $V_{\text{AHE}}\sim$ 0.3 $\mu$V, considering a Hall bias of 2 mA and a magnetoelectric node voltage $V_{\text{ME}}=0.3$ V. The Hall signal can be raised to $\sim$ 1 $\mu$V by increasing $V_{\text{app}}$ to 1 V, and further enhanced by applying a larger Hall bias. However, doing so would negatively impact the energy consumed in the read operation. Sensing such a low $\mu$V-range AH signal would require sophisticated instrumentation sense amplifiers that are area- and power-prohibitive
(e.g., 2.5 mm$^2$ area and $\sim$mW-range power~\cite{witte2008current}).

This problem can be addressed by exploring other material systems with much higher interfacial spin-orbit coupling (SOC), resulting in larger AH coefficients. 
In~\cite{zhang2014effective}, a Pt/Co/Pt tri-layer is shown to exhibit $R_{\text{s}}\sim7.3\times 10^{-10}$ $\Omega$m/T at 300 K for $t_{\text{Co}}\sim$ 10 nm, resulting in $V_{\text{AHE}}\sim$ 43.8 $\mu$V at a Hall bias of 2 mA and $V_{\text{ME}}=0.3$ V. 
Magnetic semiconductors like EuTiO$_3$ possess higher $R_{\text{s}}\sim$ $8\times 10^{-9}$ $\Omega$m/T for $t_{\text{EuTiO}_3}=$ 25 nm~\cite{takahashi2018anomalous}. However, AH signals in such samples have been detected only at very low temperatures, of 2K, at which the ME effect in Cr$_2$O$_3$ vanishes.
 
\begin{figure}[ht]
\centering
\includegraphics[scale=0.33]{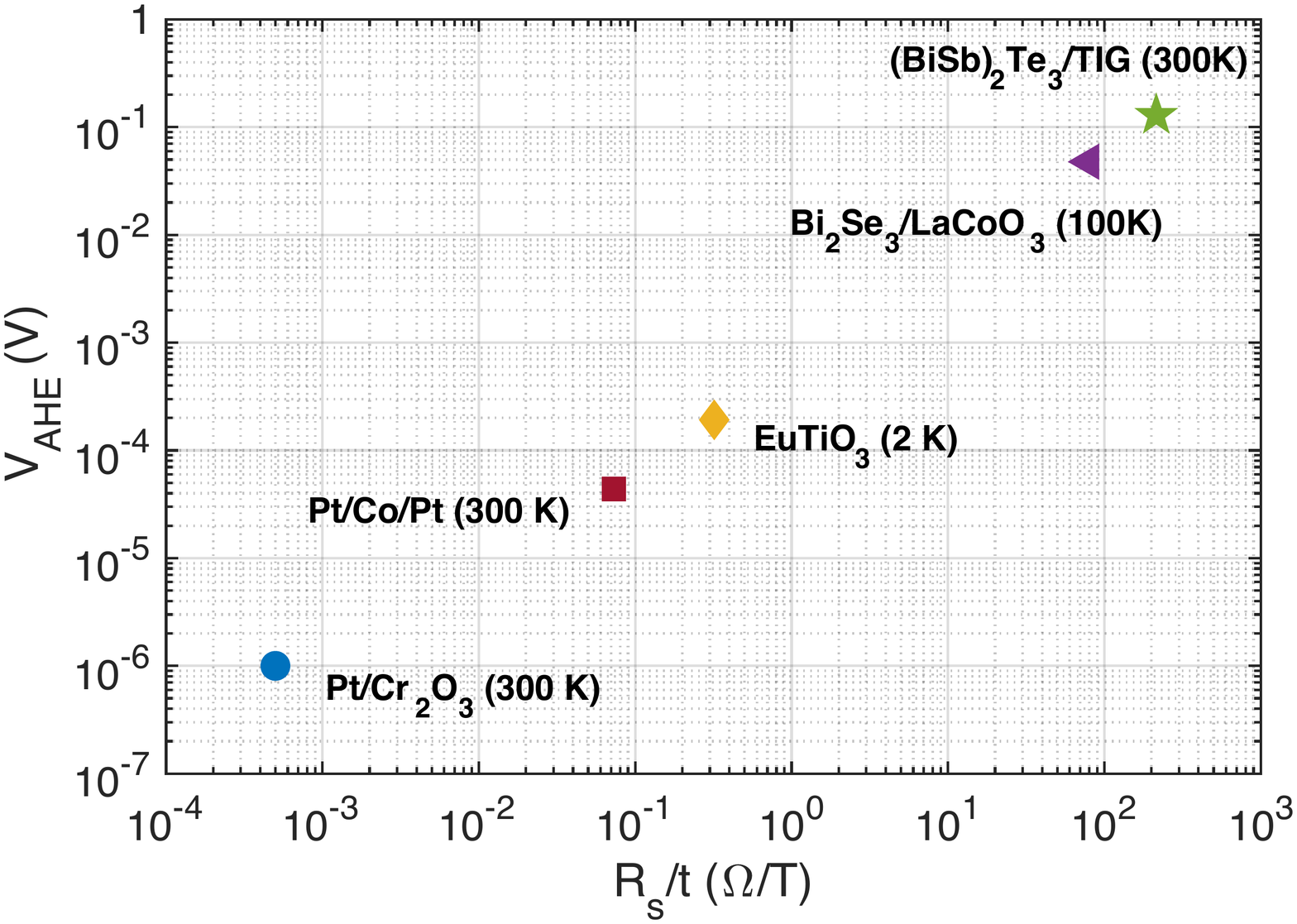}
\caption{Comparison of the AH coefficient per unit thickness and AH signal magnitude in different material systems. The AH signal $V_{\text{AHE}}$ is calculated for a Hall bias of 2 mA and a magnetoelectric node voltage $V_{\text{ME}}\sim$ 0.3 V. TIs with high interfacial SOC exhibit greater AH coefficients and can generate large AH signals, capable of being detected by conventional current sense amplifiers.}
\label{fig:R_AHE}
\end{figure}

The Hall signal could be improved in a topological insulators (TI) due to the presence of high SOC-enhanced surface states. 
For example, the Bi$_2$Se$_3$/LaCoO$_3$ stack considered in~\cite{zhu2018proximity} demonstrates $R_{\text{s}}$ as high as $\sim1.59$ $\mu\Omega$m/T at 100 K for $t_{\text{Bi}_2\text{Se}_3}\sim$ 20 nm. 
This results in a substantial improvement in the AH signal generated (i.e., $\sim47.7$ mV). 
The AH effect in the Bi$_2$Se$_3$/LaCoO$_3$ interface is ascribed to the exchange coupling between the Bi$_2$Se$_3$ layer and the ferromagnetic LaCoO$_3$ layer via the proximity effect, and is enhanced by the high interfacial SOC. 
Similarly, the (BiSb)$_2$Te$_3$/TIG system considered in~\cite{tang2017above} achieves a mV-range AH signal, though much closer to room temperature.
A comparison of $R_{\text{s}}/t$ in various material systems is illustrated in Fig.~\ref{fig:R_AHE}. 
As can be inferred, TIs are an ideal material candidate to implement the AH read-out layer with Cr$_2$O$_3$ due to the potential of a $\sim$mV-range AH signal, which can be easily read-out using a normal current latch sense amplifier~\cite{kobayashi1993current}, i.e., without the need for sophisticated sensing equipment.

\begin{figure}[ht]
\centering
\includegraphics[scale=0.37]{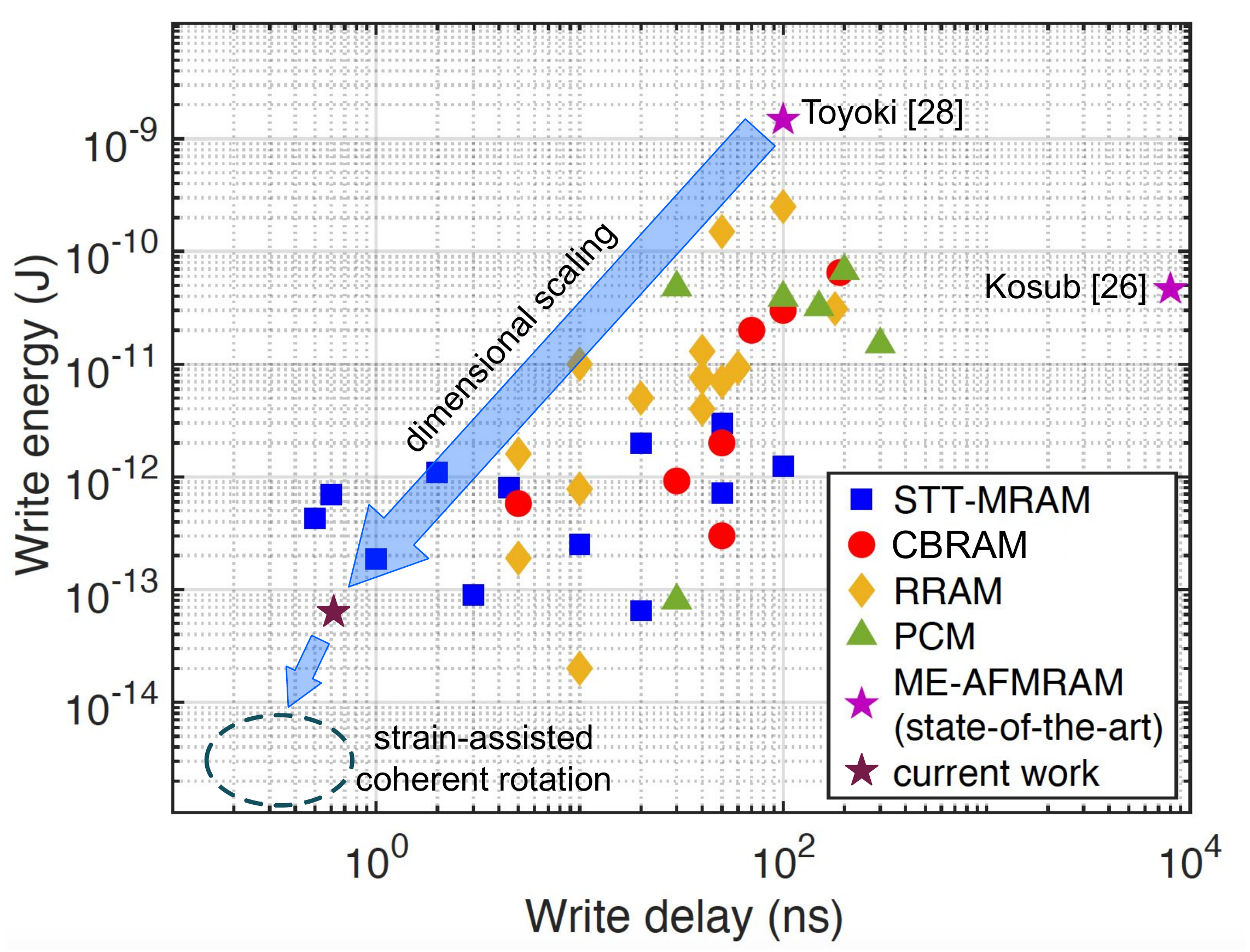}
\caption{Benchmarking the ME-AFMRAM cell considered in this work against current state-of-the-art ME-AFMRAM technology, and trends in other emerging non-volatile storage devices from~\cite{wong2016stanford}. Some important data points in this plot, representing the advances in various NVMs, include~\cite{jan2012high,gajek2012spin,liu2010ultrafast} for STT-MRAM, \cite{aratani2007novel,lin2010novel,vianello2012sb} for CBRAM, \cite{zhao2014ultrathin,sekar2014technology,goux2014role} for RRAM, and~\cite{matsui2006ta2o5,kim2010high,xiong2013self} for PCM, respectively.
The future potential of ME-AFMRAM lies in achieving ultra-fast, coherent rotation-based reversal (sub-100 ps write delay and fJ write energy) through a combination of dimensional scaling and strain-augmentation.}
\label{fig:cell_comparison}
\end{figure}

\subsubsection{Coherent rotation-based reversal}
The $\sim$ns-range write latency of the ME-AFMRAM cell can be improved drastically if the chromia order can be switched through coherent rotation. In this case, the entire chromia sample undergoes reversal homogeneously, rather than following the incoherent DW propagation. For $\mathcal{F}_{d} > 4\mathcal{K}$, the order parameter switches via damping of gyromagnetic precessions~\cite{parthasarathy2019dynamics}. However, if $\mathcal{F}_{d} < 4\mathcal{K}$, magnetization could switch due to thermal activation.
Here, the switching time is exponentially dependent on the energy barrier of the sample.
In any case,
it is thermal activation that leads to retention errors.

To realize coherent rotation in chromia, the applied ME pressure must exceed $4\mathcal{K} = 2.92\times 10^4$ J/m$^3$. For a magnetic field of 0.5 T and $\alpha_\mathrm{ME} = 3.1$ ps/m, the electric field required for coherent rotation is $1.18 \times 10^{10}$ V/m. Unfortunately, such a high electric field could lead to dielectric breakdown of chromia, given that the breakdown strength of chromia is $\sim 2\times 10^8$ V/m~\cite{sun2017local}.
A potential solution to this challenge is to reduce the effective anisotropy of the sample such that the required threshold electric field scales down. This can be achieved through a variety of techniques, including substitutional alloying and the application of mechanical strain~\cite{mu2019influence}. It is estimated that the write latency of a strain-augmented ME-AFMRAM cell can reach as low as a few 10's of ps. A comparison of the current state-of-the-art in ME-AFMRAM technology and its future potential versus trends in other emerging storage devices is presented in Fig.~\ref{fig:cell_comparison}.

\subsubsection{Material and geometrical parameters of the chromia ME-AFMRAM cell}

The simulation parameters used in our SPICE models for the chromia ME-AFMRAM are listed in the following Table~\ref{tab:params}.

\begin{table}[ht]
\centering
\footnotesize
\setlength{\tabcolsep}{1mm}
\renewcommand{\arraystretch}{1.3}
\begin{tabular}{*{3}{c}}
\hline
\textbf{Parameter} & \textbf{Value} & \textbf{Ref.}  \\
\hline
Saturation magnetization of Cr$_2$O$_3$, $M_{\text{s}}$ & $2.6\times 10^5$ A/m & \cite{artman1965magnetic} \\ \hline
Magnetoelectric coefficient of Cr$_2$O$_3$, $\alpha_{\text{ME}}$ & $3.1\times 10^{-12}$ s/m & \cite{hehl2008relativistic}\\ \hline
Uniaxial anisotropy energy of Cr$_2$O$_3$, $\mathcal{K}$ & $7300$ J/m$^3$  & \cite{foner1963high} \\ \hline
Gilbert damping constant of Cr$_2$O$_3$, $\alpha_{\text{G}}$ & $2\times 10^{-4}$  & \cite{belashchenko2016magnetoelectric}\\ \hline
Threshold ME pressure to depin DW, $\mathcal{F}_{\text{d}}$ & $25$ J/m$^3$ & \cite{parthasarathy2019dynamics} \\ \hline
Applied magnetic field, $H_{\text{app}}$ & $0.5$ T & \\ \hline
Applied voltage, $V_{\text{G}}$ & $0.3$ V & \\ \hline
Length of cell, $l$ & $60$ nm & \\ \hline
Width of cell, $w$ & $60$ nm & \\ \hline
Thickness of cell, $t$ & $10$ nm & \\ \hline
Temperature, $T$ & $292$ K & \\ \hline
$\tau_\mathrm{creep}$ (@ $\mathcal{F} = 0$) & $\sim1$ ms & \\ \hline
$\tau_\mathrm{flow}$ (@ $\mathcal{F} = 74.2$ J/m$^3$) & $\sim0.22$ ns & \\ \hline
\end{tabular}
\caption{Simulation parameters considered for the ME-AFMRAM cell.}
\label{tab:params}
\end{table}

\begin{figure*}[ht]
\centering
\includegraphics[scale=0.6]{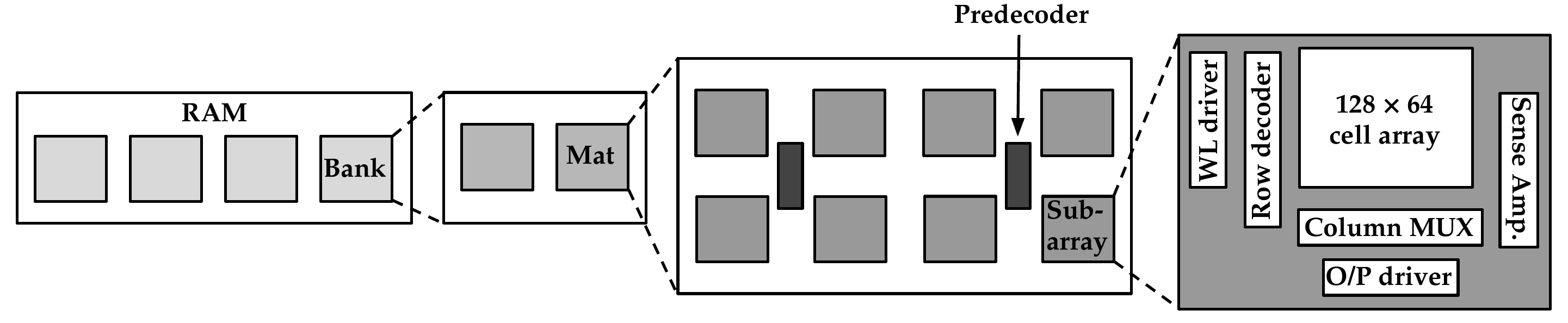}
\caption{64KB ME-AFMRAM organization with 4$\times$1 banks, 2$\times$1 mats, 4$\times$2 sub-arrays, and 128$\times$64 bit cell arrays. Here, the word length is 128 bit. The memory organization is leveraged from~\cite{dong2012nvsim}.}
\label{fig:MERAM_organization}
\end{figure*}

\begin{figure}[ht]
\centering
\includegraphics[width=\columnwidth]{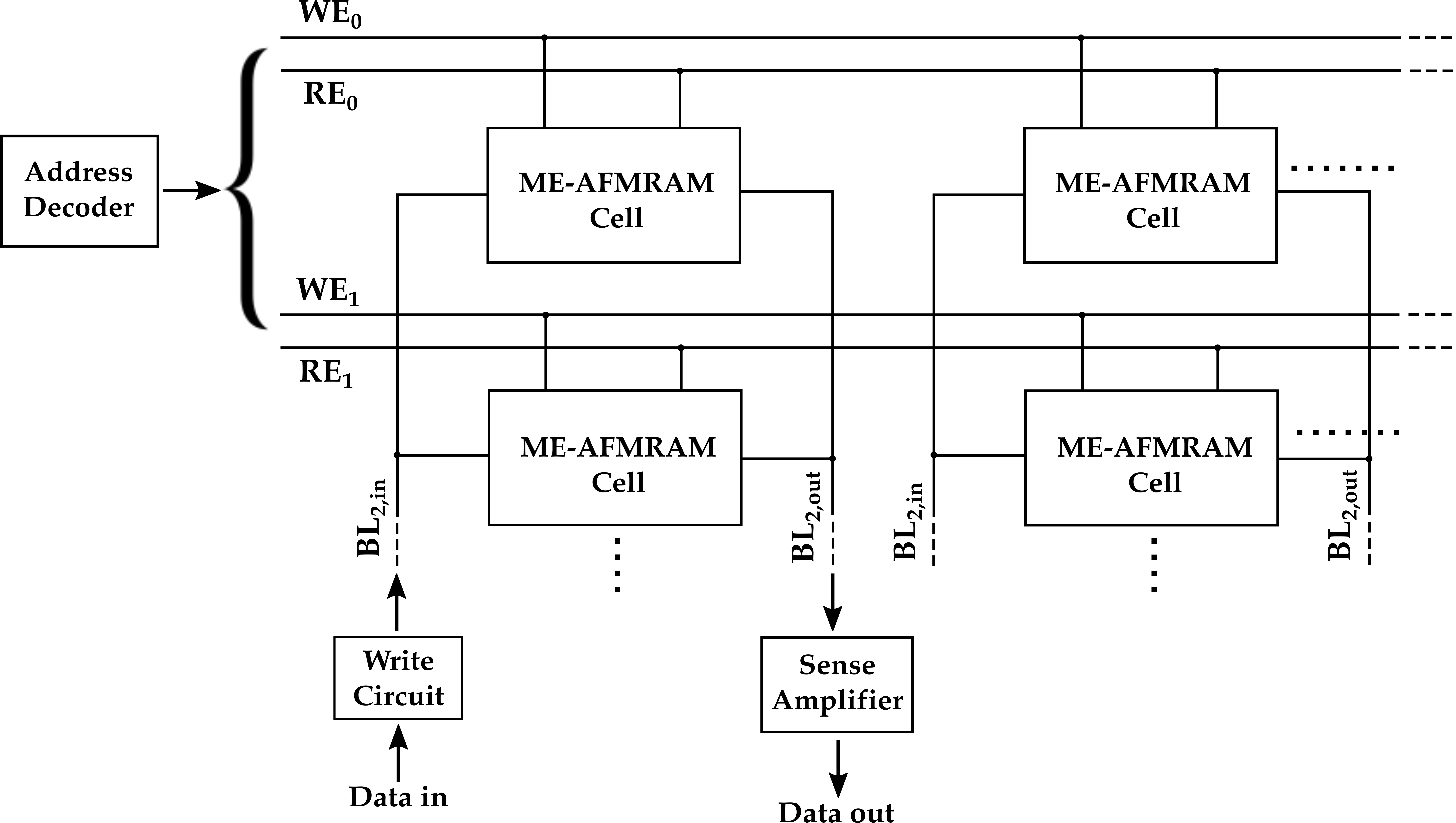}
\caption{Construction of the ME-AFMRAM cell array used in the memory architecture.
The signals BL$_{\text{i,in}}$ serve to write data into the cells when Write Enable (WE) is on, and signals BL$_{\text{i,out}}$ serve to read data from the cells when Read Enable (RE) is on.}
\label{fig:MERAM_array}
\end{figure}

\begin{figure}[ht]
\centering
\includegraphics[scale=0.43]{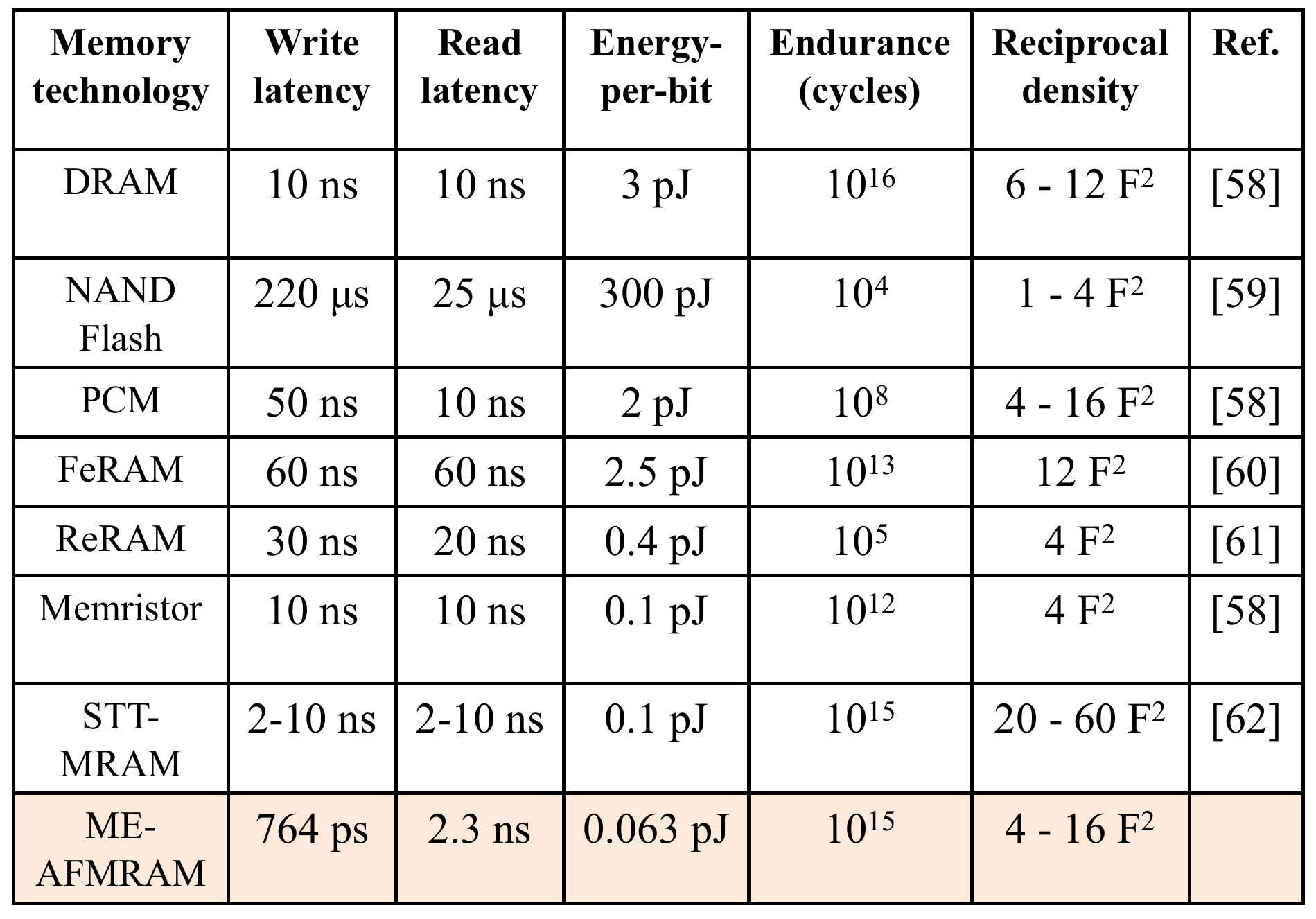}
\captionof{table}{Performance comparison of various memory technologies, from~\cite{yang2013memristive, micron_nandflash, everspin_feram, chang2016resistance, kent2015new}. 
The write and read latencies for ME-AFMRAM (DW model) are quoted for a 64KB memory with a 128-bit word line, simulated using NVSim~\cite{dong2012nvsim}. 
The energy-per-bit metric is for a single bit write onto a cell.}
\label{tab:Comparison_table}
\end{figure}

\subsection{ME-AFMRAM array}

To evaluate the system-level performance of ME-AFMRAM in the context of existing memory technologies, we simulate a 64KB DW-based ME-AFMRAM chip on NVSim, a standard tool for estimating the performance metrics of emerging NVMs~\cite{dong2012nvsim}. 
The organization of this 64KB memory, as leveraged from~\cite{dong2012nvsim}, is shown in Fig.~\ref{fig:MERAM_organization}. 
The internal architecture of the ME-AFMRAM cell array, along with the peripheral decoders, drivers and sense amplifiers, constructed at the 15-nm CMOS node, is highlighted in Fig.~\ref{fig:MERAM_array}. 
The total write latency of the 64KB ME-AFMRAM, including the parasitics and peripheral latency (133.9 ps) and the dominant cell switching time ($\sim$630 ps), is obtained as 763.9 ps from NVSim~\cite{dong2012nvsim}. 
The write latency can be improved by an order of magnitude via coherent rotation of the order parameter. 
The total read latency of the chip, obtained from NVSim~\cite{dong2012nvsim}, is $\sim$2.3 ns. 
This includes contributions from the sense amplifier (1.45 ns), bit-line parasitics (3.5 ps), decoders and other peripherals ($\sim$150 ps), and the dominant AH measurement delay in the Bi$_2$Se$_3$ layer ($\sim$0.7 ns)~\cite{kikuchi2016anomalous}. 
State-of-the-art pulsed AH measurement schemes like~\cite{kikuchi2016anomalous} are capable of operating in the GHz regime.

The output bit-line sensing can be achieved using a conventional current latch amplifier if a large-SOC material such as a TI is used to generate an AH signal in the range of tens of mV.
The read/write endurance of the ME-AFMRAM is expected to be similar to that of STT-MRAM. A comparison of the performance metrics of the ME-AFMRAM with other memory technologies at the chip-level is presented in Table~\ref{tab:Comparison_table}. It can be seen that the ME-AFMRAM offers some competitive advantages over other NVMs as well as over conventional memory systems.

\section{Application as Secure Memory}
\label{sec:security}

After conducting cell- and array-level modeling and benchmarking of the chromia-based ME-AFMRAM, we continue with the implementation of the proposed SMART memory using the ME-AFMRAM.

\subsection{Threat model}

First, we discuss the threat model, defining the strengths and capabilities of attackers, as well as the objectives and consequences of a successful attack. Most but not all attack scenarios presented here are
specific to NVMs.

\begin{itemize}

\item Attackers can launch cold-boot attacks~\cite{halderman2009lest}.
During power-down, there is some latency after the power-down sequence initiates until the moment when memory contents are completely secured. An attacker might use this gap to read out memory contents. To circumvent such attacks, memory encryption is typically employed~\cite{chhabra2011nvmm,swami2018acme}.

\item Attackers could leverage properties like sensitivity to magnetic fields and temperature fluctuations to corrupt the data or induce a DoS~\cite{jang2015self}. 
They may forcibly write specific data patterns 
to memory, which accelerates aging and 
causes memory failures.

\item With access to failure analysis equipment, attackers can also resort to advanced invasive attacks. 
The majority of such attacks
target at the back-end-of-line (BEOL), approaching from the top-most metal layer, which is also referred to as front-side attacks. 
Various countermeasures have been proposed to protect the front-side, which include protective meshes, shields, and sensors~\cite{lee19_shield,weiner18}.
In any case, \textit{bus snooping} attacks are considered beyond the scope of this work.

\item Power-dissipation signatures when reading/writing `0' and `1' within the NVM can be exploited for side-channel attacks to infer the data, through techniques like differential power analysis (DPA)~\cite{kocher1999differential} and correlation power analysis (CPA)~\cite{brier2004correlation}.

\end{itemize}

\subsection{Magnetic field and temperature attacks}
\label{sec:Mag}

\begin{figure}[ht]
\centering
\subfigure[Trajectory for magnetic field-induced switching of a FM.]{%
\label{fig:STTMRAM_traj}%
\includegraphics[scale=0.18]{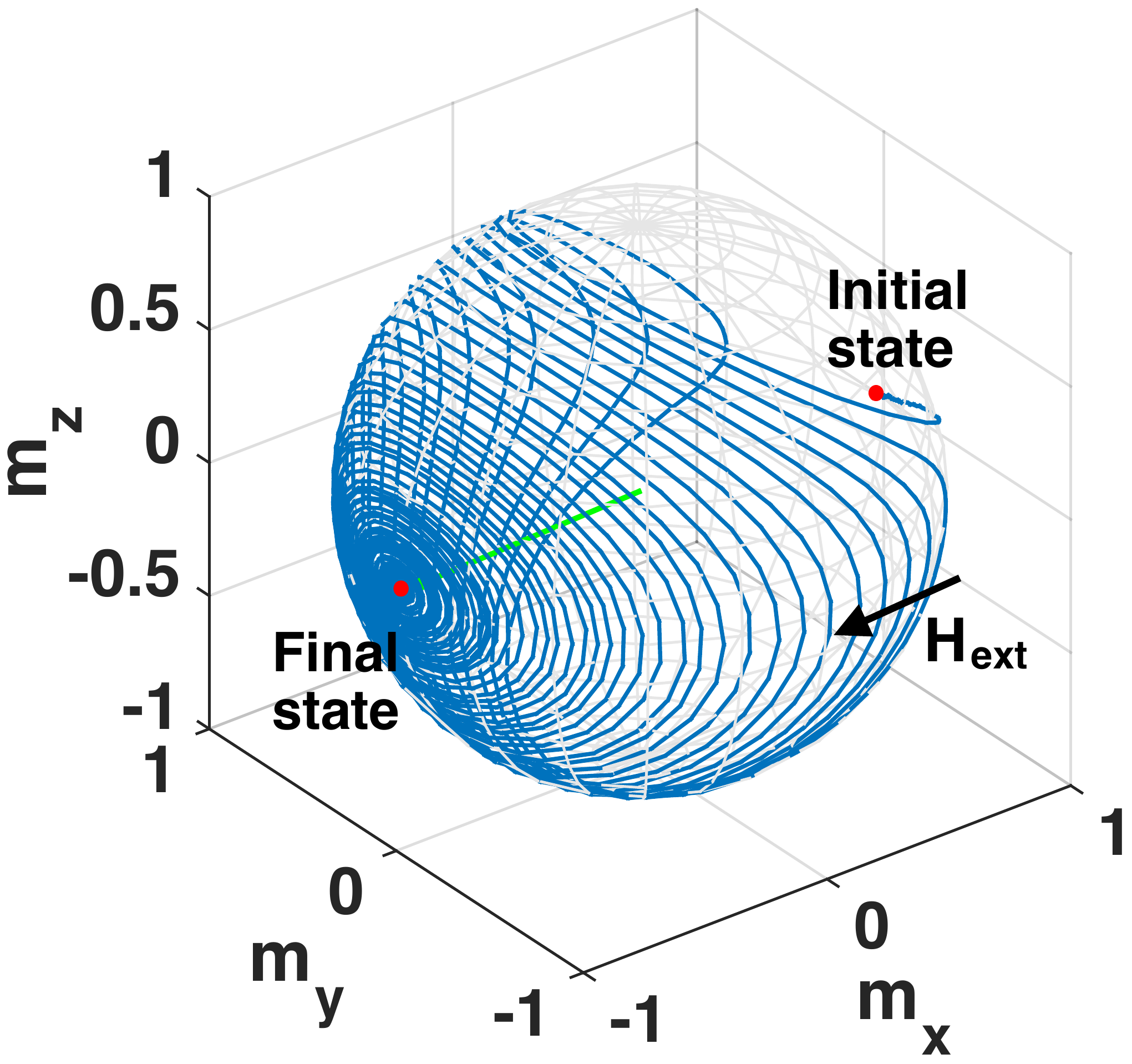}}%
\hspace{1ex}
\subfigure[Components for magnetic field-induced switching of a FM.]{%
\label{fig:STTMRAM_switching}%
\includegraphics[scale=0.23]{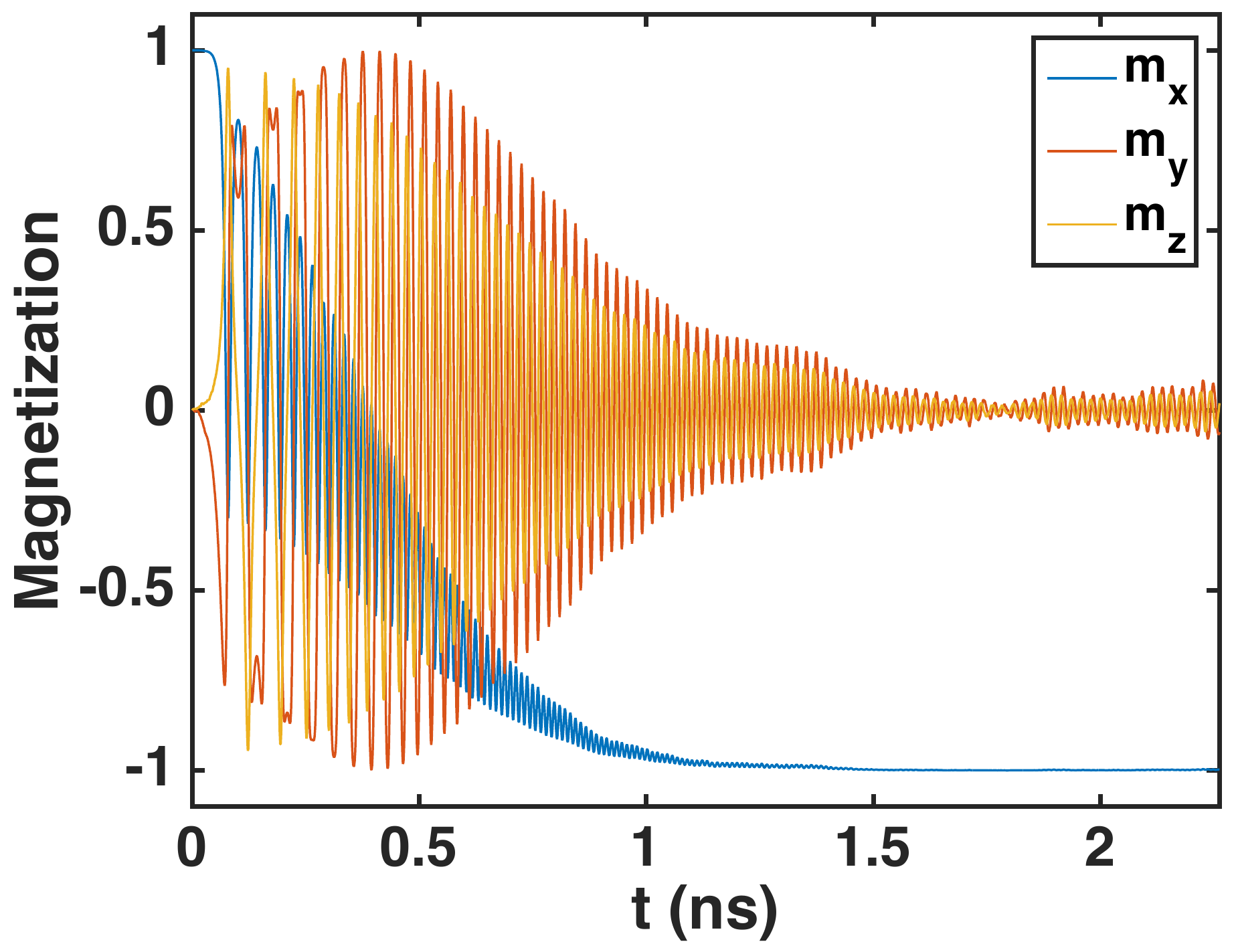}}%
\caption{The FMs in an STT-MRAM can be switched easily using external magnetic fields.}
\label{fig:STTMRAM}
\end{figure}

STT-MRAMs have FM-based MTJs as their basic building blocks. FMs possess a macroscopic magnetization (or magnetic signature) that can be probed or inferred with using an external magnetic field. 
Hence, magnetic fields can be used to infer or tamper with the stored data or even cause malfunctions in STT-MRAMs~\cite{jang2015self}.
Stray magnetic fields as small as 10 mT could cause an unintended bit flip in STT-MRAM cells. Figure~\ref{fig:STTMRAM} shows the magnetic field-induced bit flip in a representative FM, obtained by solving the Landau-Lifshitz-Gilbert equation for the FM dynamics~\cite{ament2016solving}.

\begin{figure}[ht]
\centering
\includegraphics[scale=0.3]{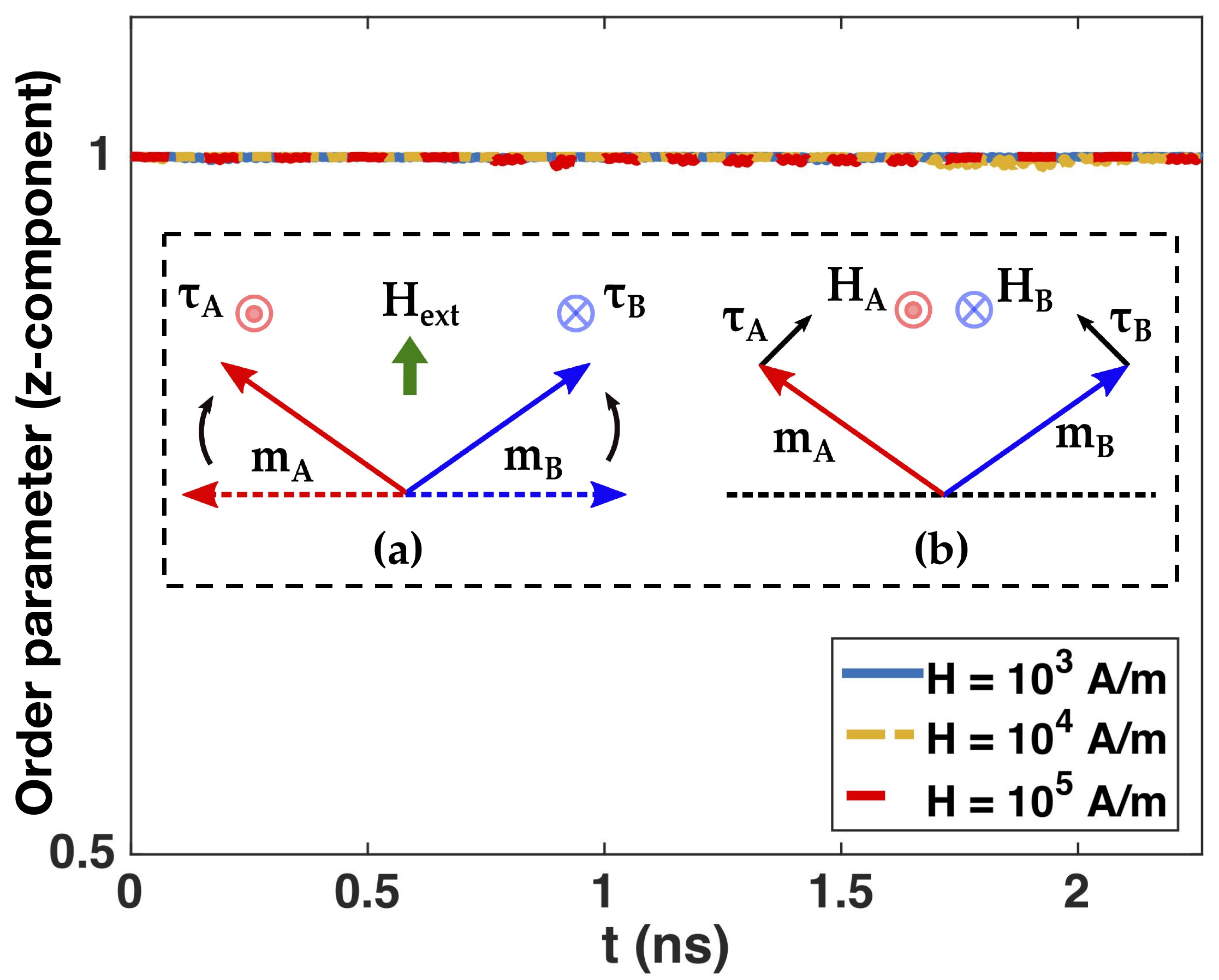}
\caption{The application of a magnetic field is unable to switch the AFM order parameter, even when increasing the field magnitude. 
Inset: (a) an external, homogeneous magnetic field may cant the sublattice moments, but it is incapable of rotating the AFM order; 
(b) staggering fields on the sublattice moments produce staggered tangential torques, which can reorient the AFM order.}
\label{fig:AFMRAM_field}
\end{figure}

AFMs, on the other hand, exhibit no external magnetic signature since their equal and opposite sublattice moments cancel each other out. 
Hence, the bulk order parameter cannot be affected by external magnetic fields. 
To switch the bulk order, staggered fields (opposite sign on opposite sublattices) must be applied on both the sublattice moments, as illustrated in Fig.~\ref{fig:AFMRAM_field} inset. 
However, an external, homogeneous magnetic field is unable to provide such a staggered field arrangement, and hence, ends up canting the sublattice moments in a way wherein the torque due to the external field is exactly balanced by the exchange torque exerted by one sublattice moment on the other~\cite{baltz2018antiferromagnetic}. 
Since external magnetic fields are unable to reorient the AFM order parameter, the SMART ME-AFMRAM is expected to be resistant to magnetic field attacks described in~\cite{jang2015self}. 
We note that switching the ME-AFM surface magnetization state using a combination of $E$ and $H$ fields would require an exact knowledge of the write cycles and the prior state of the surface, as well as
means to control the electric field explicitly, which is to be concealed from an attacker. 

With regards to temperature fluctuation-based attacks, an adversary might increase the ambient temperature of the ME-AFMRAM in an attempt to alter the stored data. 
Note that the N\'{e}el temperature of pure chromia is 308 K~\cite{shi2009magnetism}, above which the AFM ordering is destroyed. Hence, the attacker may corrupt the memory by heating it above the N\'{e}el temperature. 
To counter this, we consider Boron-doped chromia, whose N\'{e}el temperature is demonstrated experimentally to be $\sim400$ K~\cite{street2014increasing}. 
Hence, Boron-doped chromia can increase the resilience of SMART memory against temperature fluctuations. 
That is because such larger temperature fluctuations (above 400 K) are easier to detect, and countermeasures like interception of such attacks become more feasible.

\subsection{Data confidentiality attacks}\label{Encrypt}
As with all NVMs, data remanence in the SMART memory could be exploited by attackers to steal sensitive information. The most effective countermeasure against such data confidentiality attacks, including cold-boot and stolen memory-modules attacks, is to encrypt the data using a secure encryption scheme before storing it in the memory. Advanced memory encryption techniques like counter mode encryption (CME) use block ciphers such as Advanced Encryption Standard (AES) to encrypt a seed using a secret key, in order to generate a one-time pad (OTP). 
The seed for each write on a memory line consists of a secret key, the line address, and a counter value associated with that line, which is incremented with each subsequent write to the same line. Hence, the generated OTP is unique for each line address, and also for each write operation to the same address. 
The OTP is then XOR-ed with the plaintext to obtain the ciphertext, which is stored in the non-volatile main memory.
Note that the secret key used in the AES core is considered inaccessible to the attacker.

Directly applying XOR-based CME scheme to the SMART memory would result in large encryption overheads. This is because the CME scheme is tailored for NVMs like PCM and STT-MRAM, whose write time is on the order of $\sim$ns. The access latency of ME-AFMRAM is sub-ns for DW-based propagation and few 10's of ps for coherent rotation. A general encryption scheme for SMART memory, switching either via DW propagation or coherent rotation, must be such that the overall memory access latency is not adversely affected. Existing encryption solutions based on CMOS XOR gates with 10's of ps delay are rendered ineffective as their encryption time is comparable to the memory write time, resulting in idle clock cycles.

\begin{figure}[ht]
\centering
\includegraphics[scale=0.3]{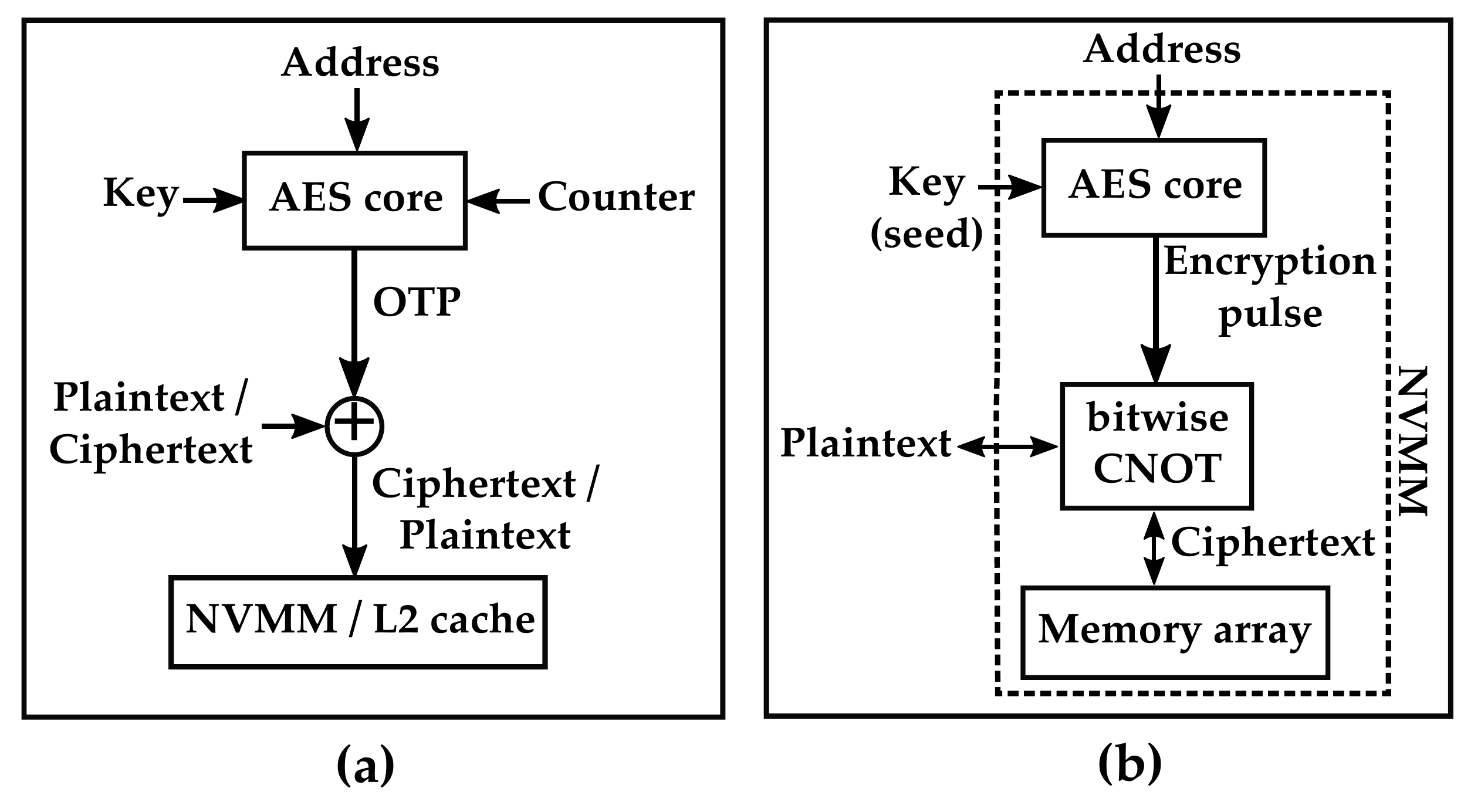}
\caption{(a) CME uses AES to generate an OTP, using the memory line address, a counter, and a secret key. The encryption and decryption is performed outside the non-volatile main memory (NVMM).
(b) \textit{Memcryption} uses a secret key and the line address as seed for AES, to generate an encryption pulse. 
That pulse is used to control the bitwise operation of CNOT gates, and is embedded in the data path within the NVMM.}
\label{fig:Memcryption_scheme}
\end{figure}

\begin{figure*}[ht]
\centering
\includegraphics[scale=0.42]{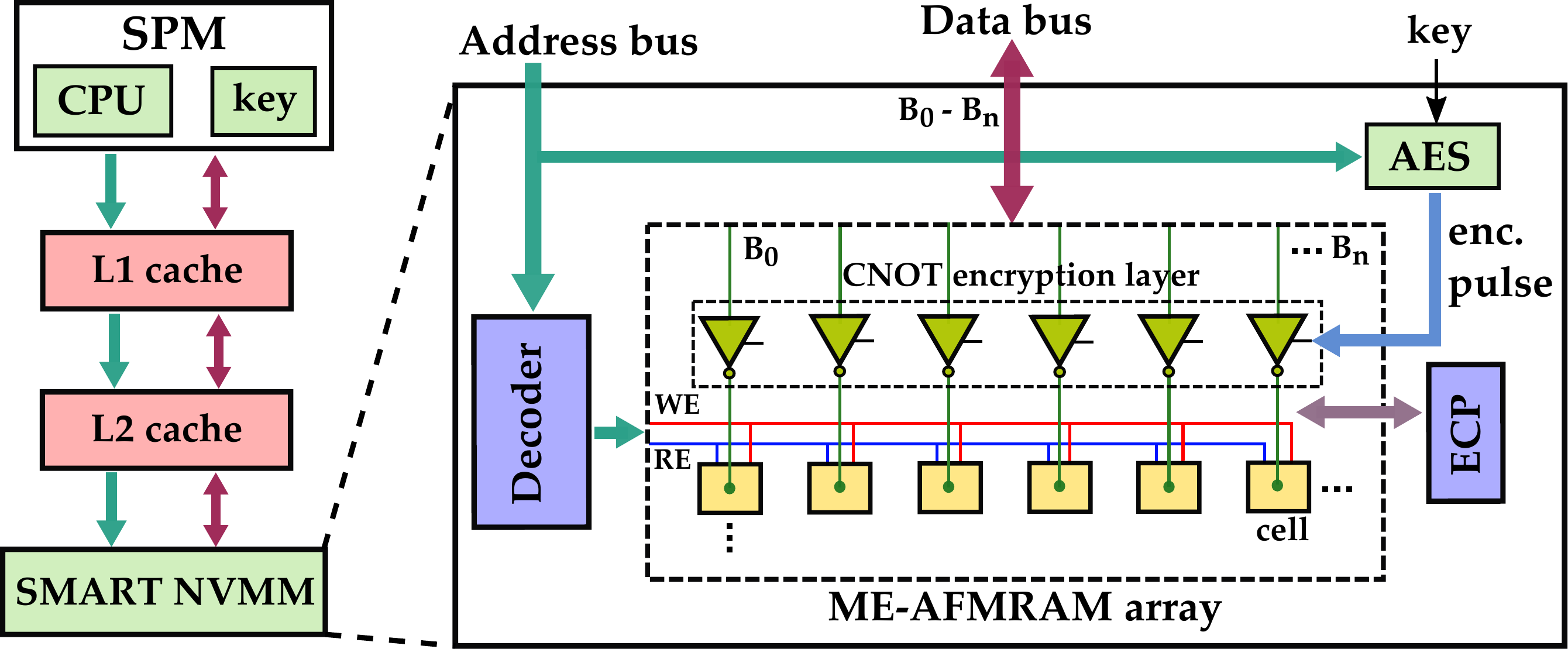}
\caption{SMART memory architecture with \textit{Memcryption}.
The CNOT layer for decryption is not shown for simplicity.}
\label{fig:Memcryption}
\end{figure*}

Here, we propose to use in-memory encryption, or \textit{Memcryption}, using bitwise CNOT (i.e., controlled-NOT) gates constructed from ME-AFM-based logic. 
By tying the encryption pulse to the control signals of CNOT gates, one can achieve such \textit{Memcryption}.
Spin devices like the ME-AFM transistor~\cite{dowben2018towards} are able to implement polymorphic logic gates, which can provide inverting or non-inverting functionality based on a control signal~\cite{patnaik2018advancing,patnaik2019spin}. 
Hence, the ME-AFM transistor is used to realize the CNOT gate. Further, the ME-AFM transistor is shown to exhibit delays as small as $\sim10$ ps, which is substantially faster than CMOS XOR gates and compatible with the SMART memory write-times.
Such homogeneity in the technology and materials by using only ME-AFM for both the memory cells and the CNOT gates will ease the fabrication. 
In \textit{Memcryption}, we embed ME-AFM transistor-based CNOT gates directly in the data path connected to the memory array; hence, the encryption is in-memory, as opposed to prior works using a separate encryption block. 
This integration of encryption and memory array is not detrimental to the memory density since ME-AFM transistors have a footprint that is substantially smaller than that of CMOS XOR gates.
Figure~\ref{fig:Memcryption_scheme} contrasts our \textit{Memcryption} scheme with prior CME techniques. 

The SMART memory architecture with \textit{Memcryption} is shown in Fig.~\ref{fig:Memcryption}.
A trusted 128-bit key, provided and stored within a secure processing module (SPM) along with the processor, is concatenated with the memory address and used as seed for AES.
The AES core, which is to be integrated on the NVM chip,\footnote{Heterogeneous spin-CMOS integration is not prohibitive since the underlying AFM technology is compatible with 
CMOS processes in the BEOL. 
In general, hybrid spin-CMOS designs have been explored in prior works~\cite{yogendra2015domain}.} thus produces an encryption pulse whose bits are used as the control bits for the CNOT gates of the in-memory encryption layer.
Depending on the control bits, the encryption layer flips bits selectively in the plaintext before performing a memory-write. 
During decryption, the same encryption pulse is generated again and used to perform bitwise CNOT operations on the ciphertext (read from memory), to obtain the plaintext.

A comparison of the \textit{Memcryption} scheme versus CME (when also applied to ME-AFMRAM) is presented in Table~\ref{overhead_comparison}. 
The array considered is a 128-bit ME-AFMRAM, while the AES and CMOS peripherals are synthesized using the 15nm \textit{NanGate} technology.
We observe that Memcryption with SMART memory 
has a better encryption latency than CME, which utilizes regular CMOS XORs.
We also note that \textit{Memcryption} helps reduce the encryption latency but is similar to CME with respect to the energy overheads. 
That is because energy dissipation is dominated by the AES core in any case.
We also reiterate that \textit{Memcryption} is tailored specifically as a memory-side scheme 
for ME-AFMRAM, to achieve low encryption latency, owing to the homogeneous delays of the memory array and the encryption layer. 
However, it may not serve well as an efficient implementation for any generic NVM.

With regards to the reliability and lifetime of the ME-AFMRAM used to construct the SMART memory, its endurance is comparable to that of STT-MRAM. 
However, it also suffers from the same errors that plague the STT-MRAM, i.e., faults in the peripheral CMOS circuitry including the access transistors~\cite{chintaluri2016analysis}. 
To address these faults and ensure the correctness of the stored data, standard error correction techniques for NVMs~\cite{swami2017reliable} like the error correction pointer (ECP) and other advanced schemes based on ECP, including ``Pay-As-You-Go''~\cite{qureshi2011pay} and ``Zombie memory''~\cite{azevedo2013zombie}, can be implemented memory-side and integrated on the ME-AFMRAM array. 
The ECP memory can be realized using homogeneous spintronics technology, including the STT-MRAM or the ME-AFMRAM itself, or by leveraging heterogeneous spin-CMOS integration.

\begin{table}[ht]
\centering
\footnotesize
\renewcommand{\arraystretch}{1.3}
\begin{tabular}{*{3}{c}}
\hline
\textbf{Encryption technique}
& \textbf{Latency} & \textbf{Energy} \\
\hline
CME~\cite{chhabra2009making} &  299.23 ps (2.99$\times$) & 17.371 pJ \\ \hline
Memcryption &  273.46 ps (2.73$\times$) & 17.370 pJ \\ \hline
\end{tabular}
\caption{Comparison for latency and energy when applying the CME and \textit{Memcryption} schemes to a 128-bit ME-AFMRAM array. The baseline latency for the unencrypted array is $\sim 100$ ps.}
\label{overhead_comparison}
\end{table}

\subsection{Power side-channel attacks}
\label{Power}

Asymmetric read/write characteristics in NVMs like STT-MRAM make them susceptible to side-channel attacks which exploit the different signatures incurred when reading/writing `1's 
and `0's bits. STT-MRAMs employ MTJs with a fixed FM reference layer, 
with another free layer either oriented parallel or anti-parallel to that reference layer. Depending on the relative orientation of these two layers,
the MTJ falls into a low or high resistance state; the low or high state corresponds to logic `0' or logic `1' state, respectively. Hence, the currents drawn for read/write operations are different depending on reading/writing a `0' or a `1'. 
Thus, an attacker could attach a resistor in a voltage-divider configuration with
the MTJ cell, monitor the voltage drops across that resistor, and perform DPA to recover the data being written to or read from the cell. In fact, such an attack was showcased against an STT-MRAM-based cache in~\cite{khan2017side}.

For the SMART memory, recall that writing is achieved using electrical fields, not currents. Further, the electric-field magnitude required for writing `0's and `1's is equivalent; see write voltage and polarization voltage traces in Fig.~\ref{fig:AFMRAM_timing}. This is because there is no reference layer or tunneling magnetoresistance in the ME-AFMRAM, which would cause asymmetricity. As for the read operation, the proximity effect-induced moment in the Pt electrode is slightly different for reading `0' or `1'. 
However, this imbalance in the Hall signals can be compensated for by introducing appropriate offsets in the Hall measurement setup, as demonstrated in~\cite{kosub2017purely}. Hence, the SMART memory can achieve symmetric signatures for both read and write and for both `0$\rightarrow$1' and `1$\rightarrow$0' 
transitions, thus thwarting any DPA-based power side-channel attacks.

\subsection{Photonic side-channel and 
backside attacks}
\label{Photonic}

Leveraging the photonic side-channel (PSC) to circumvent the security guarantees provided by cryptographic algorithms like AES and RSA has been demonstrated recently~\cite{ferrigno2008aes,schlosser2013simple}. Simple Photonic Emission Analysis (SPEA) or Differential Photonic Emission Analysis (DPEA) can be carried out using photo-emission equipment available for similar cost as that of power-analysis equipment. The essence of the PSC is to observe photo-emissions emanating for switching of CMOS transistors.
For SRAM- or DRAM-based memories, this emission can then be correlated with the data being programmed into the memory. In~\cite{ferrigno2008aes}, the PSC was found to originate when kinetic energy gained by charge carriers in the transistor channel is transferred to photons, which are visible through photo-detectors. In~\cite{schlosser2013simple}, the authors leveraged this 
information to perform a side-channel attack, ultimately recovering the full AES key. Modern-day chips use several metal layers, which interfere with the emission of photons from the frontside of any integrated circuit (IC); therefore, a natural direction is to observe the photon emission from the backside of ICs. 

While CMOS-based memory technologies like SRAM and DRAM are prone to such PSC attacks, the SMART memory is AFM-based and involves no photonic emissions emanating from transistor channels. Data read-out in the SMART memory can only be accomplished through an AH measurement setup. Further, even if an advanced attacker is able to isolate the SMART memory cell and gain access to the AH setup from the frontside, they would only be able to recover the encrypted ciphertext (as described in Sec.~\ref{Encrypt}).

\section{Conclusion}
\label{sec:conclusion}

In this paper, we present \textit{SMART: A Secure Magnetoelectric Antiferromagnet $\!$-Based Tamper-Proof Non-Volatile Memory}, by utilizing the unique properties of ME-AFMs. 
The ME-AFMRAM, which is at the core of the SMART memory, has an access latency of sub-1 ns (for DW-based switching) down to only 10's of ps (for coherent rotation switching) with an energy-per-bit of $\sim$ 0.13 pJ. 
Besides its superior performance as compared to prior NVMs like STT-MRAM and PCM, the SMART memory exhibits no sensitivity to external magnetic fields, which makes it resilient to magnetic field-based data tampering and denial of memory service attacks that commonly plague other ferromagnets-based NVMs. 
To solve the security vulnerability of data remanence (after power-down) in the SMART memory, we demonstrate a new encryption technique called \textit{Memcryption}. 
This scheme employs emerging ME-AFM-based logic to implement a CNOT-centric in-memory encryption, which is particularly tailored to reduce the encryption and decryption latency in the SMART memory. 
Further, symmetric read and write signatures for `0' and `1' bits render prominent side-channel attacks like the differential power attack futile against the SMART memory. 
Advanced photonic side-channel attacks, which are powerful threats against any CMOS IC by observing all internal transistor activity from the frontside or backside, are ineffective against the SMART memory due to the fundamentally different switching mechanism as well as the proposed \textit{Memcryption} safeguard.

\bibliographystyle{IEEEtran}
\bibliography{main}

\begin{IEEEbiography}[{\includegraphics[width=1in,height=1.25in,clip,keepaspectratio]{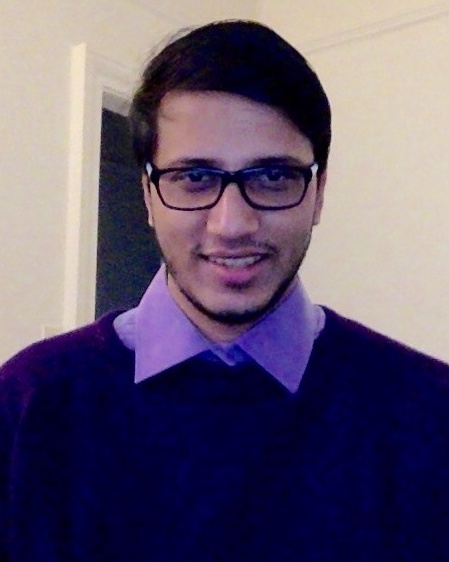}}]{Nikhil Rangarajan} (S'15--M'20)
is a Postdoctoral Associate at the Division of Engineering, New York University Abu Dhabi, UAE. 
He has Ph.D. and M.S. degrees in Electrical Engineering from New York University, NY, USA. 
His research interests include spintronics, nanoelectronics, device physics and hardware security. 
He is a member of IEEE. 
\end{IEEEbiography}

\begin{IEEEbiography}[{\includegraphics[width=1in,height=1.25in,clip,keepaspectratio]
{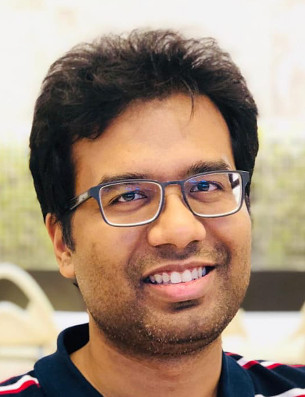}}]{Satwik Patnaik} (S'16)
received B.E.\
in Electronics and Telecommunications from the University of Pune, India and
M.Tech.\ in Computer Science and Engineering with a specialization in VLSI Design from Indian Institute of Information Technology and
Management, Gwalior, India. 
He is a final year Ph.D.\ candidate at the Department of Electrical and Computer Engineering at the 
Tandon School of Engineering with New York University, Brooklyn, 
NY, USA.
He is
also a Global Ph.D.\ Fellow with New York University Abu Dhabi, U.A.E.
He received the Bronze Medal in the Graduate category at the ACM/SIGDA Student Research Competition (SRC) held at ICCAD 2018, and the best paper award at the Applied Research Competition (ARC) held in conjunction with Cyber Security Awareness Week (CSAW), 2017.
His current research interests 
include Hardware security, Trust and reliability issues for CMOS and emerging devices 
with particular focus on low-power VLSI Design.
He is a student member of IEEE and ACM.
\end{IEEEbiography}

\begin{IEEEbiography}[{\includegraphics[width=1in,height=1.25in,clip,keepaspectratio]{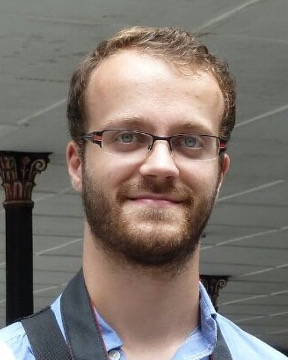}}]{Johann Knechtel}
(M'11)
received the M.Sc.\ in Information Systems Engineering (Dipl.-Ing.) in 2010 and the Ph.D.\ in Computer Engineering
(Dr.-Ing., summa cum laude) in 2014, both from TU Dresden, Germany.  
He is a Research Scientist
at the New York University, Abu Dhabi, UAE.  Dr.\ Knechtel was a Postdoctoral Researcher in 2015--16 at the Masdar Institute of Science and Technology, Abu Dhabi.  
From 2010 to 2014, he was a Ph.D.\ Scholar with the DFG Graduate School on ``Nano- and Biotechnologies for Packaging of Electronic
Systems'' hosted at the TU Dresden.  
In 2012, he was a Research Assistant with the Dept.\ of Computer Science and Engineering, Chinese University of Hong Kong, China.  
In 2010, he was a Visiting Research Student with the Dept.\ of Electrical Engineering and Computer Science, University of Michigan, USA.
His research interests cover VLSI Physical Design Automation, with particular focus on Emerging Technologies and Hardware Security.
\end{IEEEbiography}

\begin{IEEEbiography}[{\includegraphics[width=1in,height=1.25in,clip,keepaspectratio]{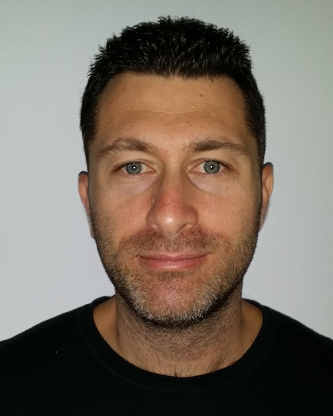}}]{Ozgur Sinanoglu} (M'11--SM'15)
is a Professor of Electrical and Computer Engineering at New York University Abu Dhabi. He earned his B.S.\ degrees, one in
Electrical and Electronics Engineering and one in Computer Engineering, both from Bogazici University, Turkey in 1999. He obtained his MS
and PhD in Computer Science and Engineering from University of California San Diego in 2001 and 2004, respectively. He has industry
experience at TI, IBM and Qualcomm, and has been with NYU Abu Dhabi since 2010. During his PhD, he won the IBM PhD fellowship award twice.
He is also the recipient of the best paper awards at IEEE VLSI Test Symposium 2011 and ACM Conference on Computer and Communication Security 2013. 

Prof.\ Sinanoglu's research interests include design-for-test, design-for-security and design-for-trust for VLSI circuits, where he has more than 180 conference and journal papers, and 20 issued and pending US Patents. 
Prof.\ Sinanoglu has given more than a dozen tutorials on hardware security and trust in leading CAD and test conferences, such as DAC, DATE, ITC, VTS, ETS, ICCD, ISQED, etc. 
He is serving as track/topic
chair or technical program committee member in about 15 conferences, and as (guest) associate editor for IEEE TIFS, IEEE TCAD, ACM JETC,
IEEE TETC, Elsevier MEJ, JETTA, and IET CDT journals.

Prof.\ Sinanoglu is the director of the Design-for-Excellence Lab at NYU Abu Dhabi. 
His recent research in hardware security and trust is being funded by US National Science Foundation, US Department of Defense, Semiconductor Research Corporation, Intel Corp and Mubadala Technology.
\end{IEEEbiography}

\begin{IEEEbiography}[{\includegraphics[width=1in,height=1.25in,clip,keepaspectratio]{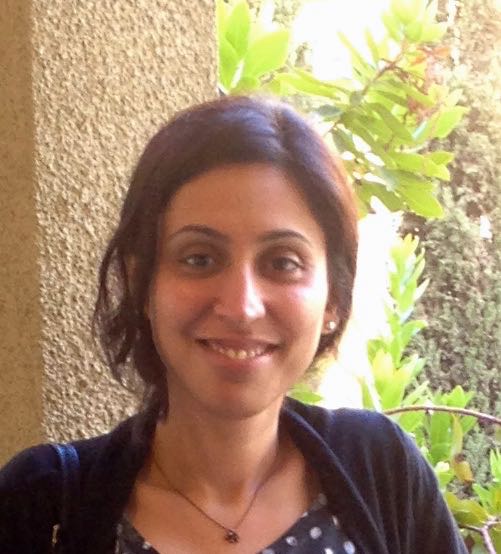}}]{Shaloo Rakheja} (M'13)
is an Assistant Professor of Electrical and Computer engineering with the Holonyak Micro and Nanotechnology Laboratory, University of Illinois at Urbana-Champaign, Urbana, IL, USA, where she works on
nanoelectronic devices and circuits. She was previously an
Assistant  Professor  of  Electrical  and  Computer  engineering with New  York  University,  Brooklyn,  NY,  USA.
Prior to joining NYU, she was a Postdoctoral Research Associate with the Microsystems Technology
Laboratories, Massachusetts Institute of Technology, Cambridge, USA. She obtained her M.S.\ and Ph.D.\ degrees in Electrical and Computer
Engineering from Georgia Institute of Technology, Atlanta, GA, USA. 
\end{IEEEbiography}

\EOD
\end{document}

%% file: settings.tex
\usepackage[utf8]{inputenc}
\usepackage[T1]{fontenc}
\usepackage{cite}
\usepackage{amsmath,amssymb,amsfonts}
\usepackage{algorithmic}
\usepackage{graphicx}
\usepackage{textcomp}
\usepackage{soul}
\usepackage{balance}

\usepackage{caption}
\DeclareCaptionFont{ieeeblue}{\color{accessblue}}
\DeclareCaptionLabelFormat{myformat}{\figcapfont{\textbf{#1} \textbf{#2}}}
\usepackage{subfigure}
\usepackage{color}

\usepackage{soul}
\newcommand{\drop}[1]{\textcolor{red}{#1}}
\renewcommand{\drop}[1]{}